\documentclass[%
reprint,
amsmath,amssymb,
aps,
]{revtex4-2}

\usepackage{graphicx}
\usepackage{dcolumn}
\usepackage[colorlinks=true, citecolor=blue, linkcolor=blue, urlcolor=blue]{hyperref}
\usepackage[mathlines]{lineno}
\usepackage{enumitem}
\newcommand{\nn}{\nonumber \\}
\newcommand{\Sec}[1]{\textit{#1.}---}
\begin{document}
	
	
	\title{Statistics of Matrix Elements of Operators in a Disorder-Free SYK model}
	
	\author{Tingfei Li}
	\email{tfli@zju.edu.cn}
	\affiliation{College of Physics Science and Technology, Hebei University, Baoding 071002, China}
	\affiliation{Hebei Key Laboratory of High-precision Computation and Application of Quantum Field Theory, Baoding, 071002, China}
	\affiliation{Hebei Research Center of the Basic Discipline for Computational Physics, Baoding, 071002, China}
	\affiliation{Kavli Institute for Theoretical Sciences (KITS), University of Chinese Academy of Sciences, Beijing 100190, China}
	\author{Shuanghong Li}%
	\email{leesh@zju.edu.cn}
	\affiliation{%
		Zhejiang Institute of Modern Physics, Zhejiang University, Hangzhou, 310027, P. R. China,
	}%

	\date{\today}%
	
	\begin{abstract}
		Recently, studies have explored the statistics of matrix elements of local operators in the Lieb-Liniger model. It was found that the probability distribution function for off-diagonal matrix elements $\langle \boldsymbol{\mu}|\mathcal{O}|\boldsymbol{\lambda} \rangle$ within the same macro-state is well described by the Fr\'{e}chet distributions. This represents a significant development for the Eigenstate Thermalization Hypothesis (ETH).
		In this paper, we investigate a similar phenomenon in another solvable model: the disorder-free Sachdev-Ye-Kitaev (SYK) model. The Hamiltonian of this model consists of 4-body interactions of Majorana fermions. Unlike the conventional SYK model, the coupling strengths in this model are fixed to a constant, earning it the name ``disorder-free.'' We evaluate the matrix elements of operators constructed from products of $n$ Majorana fermions: $\mathcal{O} = \chi_{a_1}\chi_{a_2}\ldots \chi_{a_n}$. For a general choice of indices and $n \geq 4$, we find that the statistics of the off-diagonal matrix elements are well-fitted by a generalized inverse Gaussian distribution rather than Fr\'{e}chet distributions.
	\end{abstract}
	
	\maketitle

	\Sec{Introduction}
	Fully characterizing the mechanism underlying the emergence of equilibrium statistical mechanics from the nonequilibrium evolution of many-particle quantum systems has been a long-standing challenge in theoretical physics. A key element of our current understanding is the eigenstate thermalization hypothesis (ETH) \cite{PhysRevA.43.2046,PhysRevE.50.888,Srednicki_1999,doi:10.1080/00018732.2016.1198134}, which relates thermalization in non-integrable quantum systems to the statistical properties of matrix elements of (local) operators in energy eigenstates. The traditional ETH gives
	\begin{align}
		\mathcal{O}_{nm} &= O(\bar{E})\delta_{nm} + e^{-\frac{1}{2}S(\bar{E})} f_{\mathcal{O}}(\bar{E},\omega) R_{nm}, \nonumber \\
		\bar{E} &= \frac{E_n+E_m}{2}, \quad \omega = E_n - E_m,
	\end{align}
	where $S(\bar{E})$ is the thermodynamic entropy at energy $\bar{E}$. Here $R_{nm}$ are random variables with zero mean and non-vanishing variance. $O(\bar{E})$ and $f_{\mathcal{O}}(\bar{E},\omega)$ are smooth functions of their arguments. The ETH does not predict the explicit statistics of the off-diagonal matrix elements of an operator. A recent focus has been to clarify the statistical properties of the random variables $R_{nm}$ and test ETH in integrable systems \cite{PhysRevLett.129.170603,korepin1993quantuminversescatteringmethod,Takahashi_1999,Essler_Frahm_2005,Gaudin_2014}, whose results have been restricted to very small system sizes or low particle numbers. Recently, progress has been made in studying the statistics of $R_{nm}$ in the thermodynamic limit for a solvable model (the Lieb-Liniger model) \cite{essler2023statisticsmatrixelementslocal}.

	The Lieb-Liniger model describes $N$ bosons on a circle of length $L$ interacting via a $\delta$-function potential \cite{lieb1963exact, korepin1993quantum}:
	\begin{align}
		H = \sum_{j=1}^{N} -\frac{\partial^{2}}{\partial x_{j}^{2}} + 2c \sum_{i<j} \delta(x_i - x_j).
	\end{align}
	Its second-quantized form is
	\begin{align}
		H = \int dx \left( -\Phi^\dagger(x) \partial_x^2 \Phi(x) + c \bigl(\Phi^\dagger(x)\bigr)^2 \bigl(\Phi(x)\bigr)^2 \right).
	\end{align}
	The bosonic theory can be transformed into a fermionic one \cite{Creamer-1980jn} via the mapping
	\begin{align}
		\Phi^\dagger(x) = \Psi^\dagger(x) \, e^{\text{i}\pi \int_{-\infty}^x dz \, \Psi^\dagger(z)\Psi(z)},
		\label{BoseFermi}
	\end{align}
	where $\Psi(x)$ is a complex fermion field satisfying the canonical anti-commutation relations $\{\Psi(x),\Psi^\dagger(y)\} = \delta(x-y)$.
	
	The Lieb-Liniger model is exactly solvable by the coordinate Bethe ansatz \cite{lieb1963exact, korepin1993quantum}, and its low-energy excitations are described by \textbf{fermionic} quasi-particles. In the thermodynamic limit $N,L \to \infty$ with fixed density $N/L$, one can define a macro-state at inverse temperature $\beta$. Each macro-state corresponds to a large number of micro-states, which are characterized by a set of Bethe rapidities $\{\lambda_i\}_{i=1}^N$ and denoted by $|\boldsymbol{\lambda}\rangle$.
	
	Consider the matrix element $\langle \boldsymbol{\mu} | \mathcal{O} | \boldsymbol{\lambda} \rangle$ for a local operator such as $\mathcal{O} = \Phi(0)$. A convenient quantity to study is
	\begin{align}
		\mathcal{M}_{\boldsymbol{\lambda},\boldsymbol{\mu}} = -\frac{1}{L} \ln\left[ \frac{|\langle \boldsymbol{\mu} | \Phi(0) | \boldsymbol{\lambda} \rangle|^2}{\langle \boldsymbol{\lambda} | \boldsymbol{\lambda} \rangle \langle \boldsymbol{\mu} | \boldsymbol{\mu} \rangle} \right].
	\end{align}
	A crucial feature of the model is that the operator $\Phi(0)$, although local in the bosonic description, becomes non-local in the fermionic Fock space. For an $N$-particle micro-state $\boldsymbol{\lambda}$ and an $(N-1)$-particle micro-state $\boldsymbol{\mu}$, $\mathcal{M}_{\boldsymbol{\lambda},\boldsymbol{\mu}}$ is always non-zero. Remarkably, if $\boldsymbol{\lambda}$ is fixed and $\boldsymbol{\mu}$ is chosen randomly from the same macro-state, the distribution of $\mathcal{M}_{\boldsymbol{\lambda},\boldsymbol{\mu}}$ is found to follow the \textbf{Fréchet distribution} \cite{frechet1927}:
	\begin{align}
		P_{\alpha,\beta,\nu}(x>\nu)=\frac{1}{(x-\nu)^{1+\alpha}}\exp\left[-\left(\frac{x-\nu}{\beta}\right)^{-\alpha}\right]\ ,
	\end{align}
   here $P_{\alpha,\beta,\nu}(x<\nu)=0$. 
   The authors conjecture that their results carry over to other integrable models. A natural question is whether such statistics hold for other solvable systems, particularly zero-dimensional integrable models.
   
   The Sachdev-Ye-Kitaev (SYK) model \cite{sy1993, kitaev, trunin, rosenhaus2019introduction, chowdhury2022sachdev} plays an important role in quantum chaos \cite{qchaos2018,d2016quantum, borgonovi2016quantum, liu2020quantum}, quantum gravity \cite{Garc2016, Jensen_2016}, and condensed matter physics \cite{Davison_2017, Luo_2019, Rossini_2020, Chowdhury_2022}. In this paper, we focus on a solvable variant: the disorder-free SYK model studied in \cite{Ozaki:2024wpj}. Other related models have also been investigated \cite{witten2019syk, klebanov2017uncolored, iyoda2018, krishnan2018contrasting, Balasubramanian2021, Claps,guan2026exactlysolvabledisorderfreequantum}. As the name suggests, this model replaces the random couplings $J_{ijkl}$ of the standard SYK model  with deterministic ones. The disorder-free SYK model in \cite{Ozaki:2024wpj} exhibits weak quantum chaos, as diagnosed by out-of-time-order correlators (OTOCs) \cite{larkin1969, Polchinski2016, Maldacena2016prd, Dowling2023}. Unlike the Lieb-Liniger model, this system is zero-dimensional, consisting of all-to-all interactions; consequently, there is no notion of a ``local" operator in the spatial sense. In this paper, we investigate the statistics of matrix elements of operators constructed from products of Majorana fermions.

	\Sec{The model}
	We consider the model introduced in \cite{Ozaki:2024wpj}. The original Hamiltonian is
	\begin{align}
		H_{4}=\text{i}^{2}\sum_{1\le j_{1}<j_{2}<j_{3}<j_{4}\le N}\chi_{j_{1}}\chi_{j_{2}}\chi_{j_{3}}\chi_{j_{4}},
	\end{align}
	where $\text{i}=\sqrt{-1},\{\chi_i,\chi_j\}=2\delta_{ij}$. 
	For simplicity, in this paper we set \(\hbar\) and the Boltzmann constant \(k_B\) equal to 1.
	Introducing a quadratic Hamiltonian
	\begin{align}
		H_{2}=\text{i}\sum_{1\le j_{1}<j_{2}\le N}\chi_{j_{1}}\chi_{j_{2}}
	\end{align}
	and the transformation
	\begin{align}
		f_{k}^{\pm}=\frac{1}{\sqrt{2N}}\sum_{j=1}^{N}e^{\mp \text{i}\left(j-1\right)\theta_{k}}\chi_{j},\qquad \theta_{k}=\frac{\left(2k-1\right)\pi}{N},
	\end{align}
	we find that $f^{\pm}_k$ are complex fermion creation and annihilation operators satisfying $\{f^+_i,f^-_j\}=\delta_{ij}$, and
	\begin{equation}
		\chi_j=\sqrt{\frac{2}{N}}\sum_{s=\pm}\sum_{k=1}^{N/2}e^{\text{i}s(j-1)\theta_k }f^s_k.
		\label{chi_f}
	\end{equation}
	The spectra are then given by
	\begin{align}
		H_{2}&=\sum_{k=1}^{N/2}\epsilon_{k}\left(n_{k}-\frac{1}{2}\right),\qquad \epsilon_{k}=2\cot\frac{\theta_{k}}{2}, \nonumber \\
		H_4&=\frac{1}{2}H_{2}^{2}+E_{0},\qquad E_{0}=-\frac{N(N-1)}{4},
	\end{align}
	where $n_k\equiv f_k^+f_k^-$ takes values $\{0,1\}$. For convenience, we set $N_F\equiv N/2$.
	
	In this paper, we consider the model defined by the Hamiltonian
	\begin{align}\label{eq:H}
		H=\frac{1}{N_F}H_4+ C H_2,
	\end{align}
	where the constant $C$ is chosen as
	\begin{align}
		C=\sum_{j=1}^{N/2}\cot\frac{\left(j-1/2\right)\pi}{N}.
	\end{align}
	This choice of $C$ simplifies the expressions; setting $C=0$ does not alter the results. 
	
	\Sec{Microstate}
	Two Majorana fermions $\chi$ combine to form a complex fermion, giving rise to $N_F = N/2$ single-particle energy levels. To construct the thermal state, we must specify the microstates. These are labeled by the quantum numbers $k = 1, \dots, N_F$ associated with the rapidities $\theta_k$.
	
	A microstate is represented by an occupation number list
	\begin{align}
		|\boldsymbol{n}\rangle = |n_1, n_2, \ldots, n_{N_F}\rangle,
	\end{align}
	where $n_i \in \{0,1\}$. In the thermodynamic limit $N_F \to \infty$, the discrete set $\{\theta_k\}$ becomes a continuous variable $\theta \in [0,\pi]$. A macrostate is then characterized by the density of states $\rho(\theta)$, defined such that
	\begin{align}
		N_F \rho(\theta) \Delta \theta = \text{number of particles in } [\theta, \theta+\Delta\theta].
	\end{align}
	
\begin{figure}[!t]
	\centering
	\includegraphics[width=0.9\linewidth]{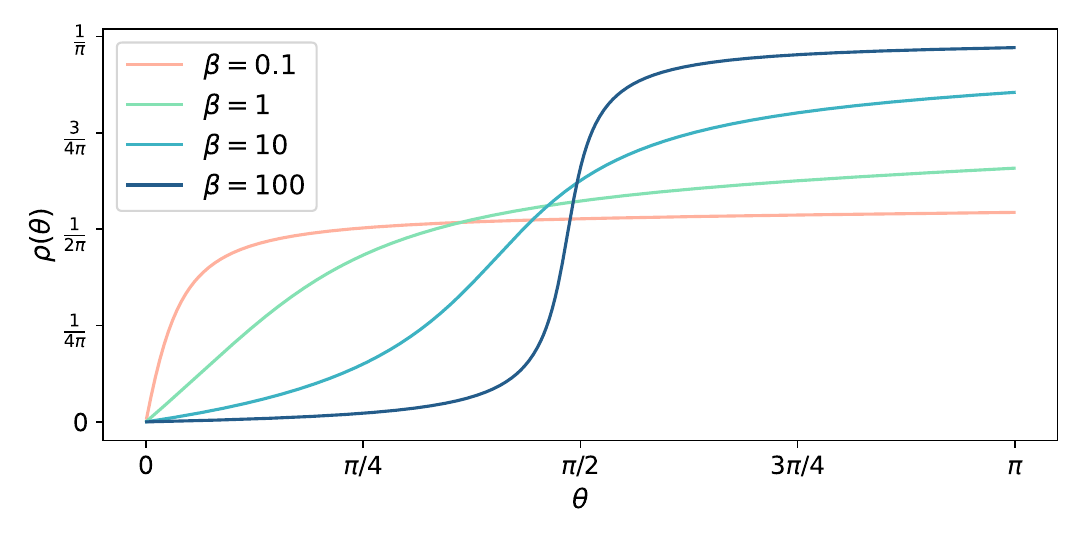}
	\caption{$\rho(\theta)$ under different temperatures; $\lambda=1/2$ here.}
	\label{rho_theta}
\end{figure}

	For a given occupation ratio $0 \le \lambda \le 1$ and inverse temperature $\beta$, the macrostate is determined by the coupled integral equations
	\begin{align}
		\int_0^\pi \rho(\theta) \, d\theta &= \lambda, \\
		\frac{1}{\rho(\theta)} - \frac{1}{\rho_h(\theta)} &= \alpha + \beta \epsilon(\theta) \int_0^\pi \epsilon(s) \rho(s) \, ds,
	\end{align}
	where $\rho(\theta)$ is obtained numerically. The solution is discussed in detail in Appendix~\ref{state_of_eq}, and typical profiles of $\rho(\theta)$ are shown in Fig.~\ref{rho_theta}.

	\Sec{Results}
	We define the product of Majorana fermions for an ordered list $\{a\}$ of length $|a| = k$ as
	\begin{align}
		\chi_{\{a\}} \equiv \chi_{a_1}\chi_{a_2}\cdots \chi_{a_k}, \qquad a_i < a_j \;\text{for all}\; i<j,
	\end{align}
	so that any operator can be expanded in the basis of these products:
	\begin{align}
		\mathcal{O} = \sum_{\{a\}} c_{\{a\}} \chi_{\{a\}}.
	\end{align}
	In this paper we focus on the statistics of the matrix elements
	\begin{align}
		\langle \boldsymbol{n} | \chi_{\{a\}} | \boldsymbol{m} \rangle.
	\end{align}
	In analogy with the Lieb–Liniger model, we introduce a similar quantity
	\begin{equation}
		\mathcal{M}_{\{a\}} = -\frac{1}{|a|}\ln\bigl|\langle \boldsymbol{n} | \chi_{\{a\}} | \boldsymbol{m} \rangle\bigr|^2 + \ln\frac{2}{N},
	\end{equation}
	where the constant $\ln\frac{2}{N}$ is added to reduce the dependence on $N$. We fix a reference state $|\boldsymbol{m}\rangle$ and sample $|\boldsymbol{n}\rangle$ from the same macrostate.
	
\begin{figure}[!t]
		\centering
		\includegraphics[width=0.925\linewidth]{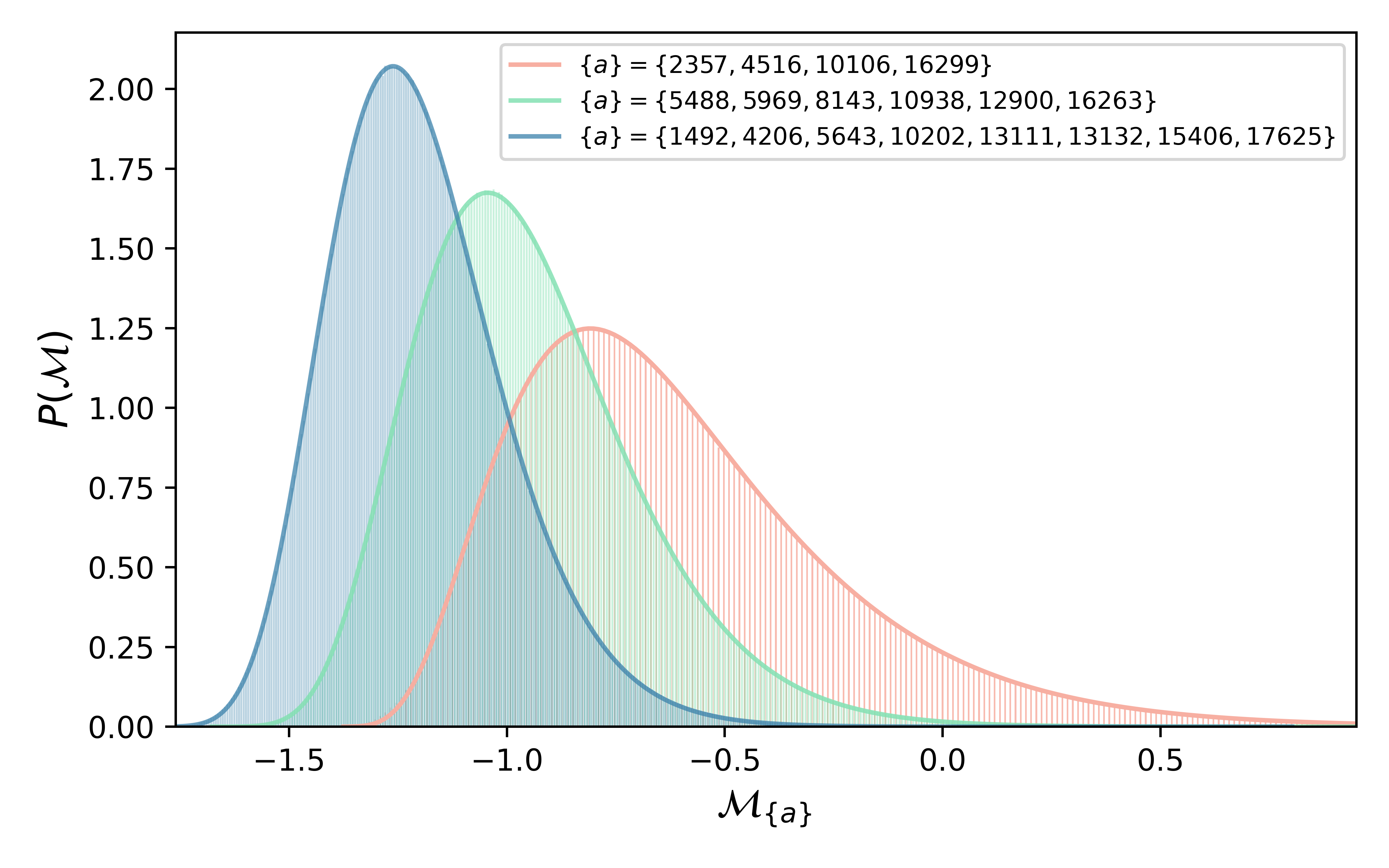}
		\includegraphics[width=0.925\linewidth]{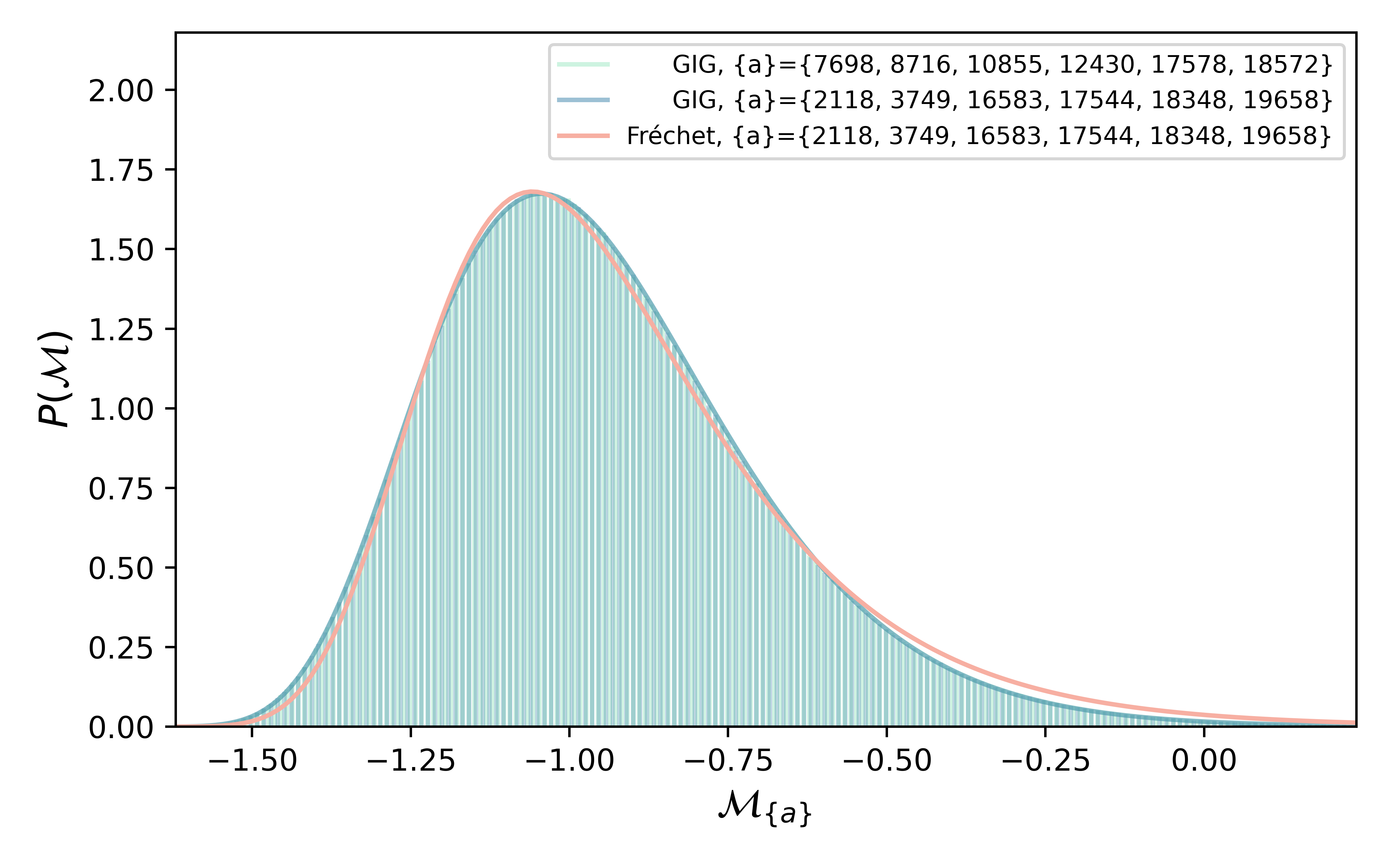}
		\caption{Distribution of $\mathcal{M}_{\{a\}}$ ($\beta=1$, $\lambda=1/2$, $N=20000$; $10^7$ samples). Upper: Histograms for three randomly chosen index sets $\{a\}$ of sizes $|a|=4$, $6$, and $8$, overlaid by generalized inverse Gaussian (GIG, $p=1$) curves. Lower: Comparison between GIG and Fréchet fits for $|a|=6$: the (GIG, $p=1$) fit is accurate, while the Fréchet fit is worse. Meanwhile, two different $\{a\}$ sets (green and blue) yield indistinguishable distributions, with their respective histograms and fitted curves overlapping.}
		\label{frechet_gig}
	\end{figure}

	The main result of this paper is the distribution of $\mathcal{M}_{\{a\}}$. Because the model is all-to-all, one might expect this distribution to depend only on the size $|a|$ and not on the specific indices chosen; we will later show that this property indeed holds in general. For a generic choice of $\{a\}$ with $|a|\ge 4$, we find that the distribution of $\mathcal{M}_{\{a\}}$ is well described by the \textbf{generalized inverse Gaussian distribution} (GIG) \cite{wiki:generalized_inverse_gaussian_distribution} with parameter $p=1$, rather than by a Fréchet distribution. The GIG density is nonzero for $x>\nu$
	\begin{align*}
		P_{p,\eta,\sigma,\nu}(x>\nu)=\frac{(\eta/\sigma)^{p/2}(x-\nu)^{p-1}}{2K_{p}\bigl(2\sqrt{\eta\sigma}\bigr)}\,e^{-\eta(x-\nu)-\frac{\sigma}{x-\nu}}
		\label{gig}
	\end{align*}
	and for $p=1$ its log‑density is a hyperbola $-\eta(x-\nu)-\sigma/(x-\nu)$. As shown in Fig.~\ref{frechet_gig}, the GIG provides a much better fit than the Fréchet distribution for large $|a|$; we have verified this for $|a|\lesssim30$.
	
	
   \Sec{Method}
	An efficient method is needed to obtain the numeric value of $\langle \boldsymbol{n} |\chi_{\{a\}} |\boldsymbol{m} \rangle$. Consider the states $|\boldsymbol{m}\rangle=f_{p_1}^\dagger \cdots f_{p_m}^\dagger|\Omega\rangle$ and $|\boldsymbol{n}\rangle=f_{q_1}^\dagger \cdots f_{q_n}^\dagger|\Omega\rangle$. Let $\boldsymbol{M}=\{p_1,\cdots,p_m\}$ and $\boldsymbol{N}=\{q_1,\cdots,q_n\}$ be their respective index sets, and let their relative complements be $\{u_1, \cdots, u_{\bar{m}}\}\equiv \boldsymbol{M}-\boldsymbol{M}\cap\boldsymbol{N}$ and $\{v_1, \cdots, v_{\bar{n}}\}\equiv\boldsymbol{N}-\boldsymbol{M}\cap\boldsymbol{N}$. The difference between $|\boldsymbol{n}\rangle$ and $|\boldsymbol{m}\rangle$ is therefore characterized by the Hamming distance~\cite{wiki:hamming_distance} $d_{\boldsymbol{nm}} =\bar{n}+\bar{m}$. For $\langle \boldsymbol{n} |\chi_{\{a\}} |\boldsymbol{m} \rangle$ to be non-zero, we must have $\chi_{\{a\}} |\boldsymbol{m} \rangle=|\boldsymbol{n}\rangle$, up to an overall factor. To ensure that the differences between $|\boldsymbol{m}\rangle$ and $|\boldsymbol{n}\rangle$ are eliminated by $\chi_{\{a\}}$, the distance must satisfy $d_{\boldsymbol{nm}} \leq |a|$. Additionally, $|a|-d_{\boldsymbol{nm}}$ must be even: if an $f^\dagger_k$ with $k\notin \boldsymbol{N}$ acts on the ket, an $f_k$ must also act on the ket to eliminate the element not in $\boldsymbol{N}$. In the limit $N_F\to\infty$, the probability of each allowed distance is proportional to $\binom{N_F}{d_{\boldsymbol{nm}}}$, so the configuration with $d_{\boldsymbol{nm}} = |a|$ dominates.

	When $d_{\boldsymbol{nm}} = |a|$, for the $\chi$ given in Eq.~\eqref{chi_f}, only the terms $\propto \langle \boldsymbol{n}|f^\dagger_{v_1}\cdots f^\dagger_{v_{\bar{n}}}\, f_{u_1}\cdots f_{u_{\bar{m}}}|\boldsymbol{m} \rangle$ are nonvanishing in the expansion of $\langle\boldsymbol{n}|\chi_{a_1}\cdots\chi_{a_{|a|}}|\boldsymbol{m}\rangle$. Summing up all such terms, by anticommutativity of fermions, we can write
	\begin{equation}
		\langle \boldsymbol{n} |\chi_{\{a\}} |\boldsymbol{m} \rangle =(-1)^P \text{det}(\mathcal{A})\ ,
		\label{det_A}
	\end{equation}
	where the matrix elements are given by
	\begin{equation}
		\mathcal{A}_{ij}=\sqrt{\frac{2}{N}}\, \text{exp}[\text{i}\, s_i(a_j-1)\theta_{k_i}],\quad 1\leq i,j \leq |a|
		\label{det_A_kl}
	\end{equation}
	with $s_i=+1$ for $k_i\in\{v_1, \cdots, v_{\bar{n}}\}$ and $s_i=-1$ for $k_i\in\{u_1, \cdots, u_{\bar{m}}\}$. The overall sign $(-1)^P$ depends on the difference between $|\boldsymbol{m} \rangle$ and $|\boldsymbol{n}\rangle$; see Appendix~\ref{n_c_m} for detail.
	
\begin{figure}[!t]
		\raggedright
		\includegraphics[width=0.95\linewidth]{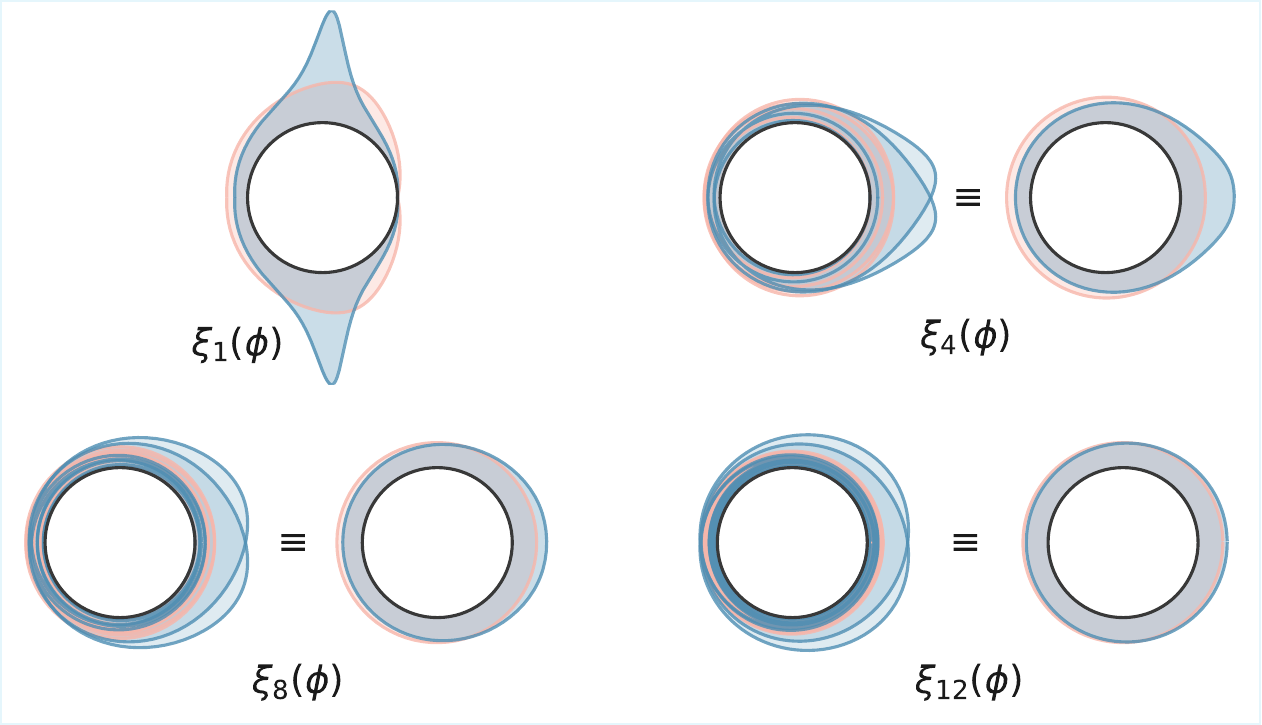}
		\caption{Schematic illustrating why $|\langle \boldsymbol{n} |\chi_{\{a\}} |\boldsymbol{m} \rangle|$ is generally insensitive to temperature and the index set $\{a\}$. Here, $\beta=10$ (red) and $100$ (blue); the scales are illustrative. Different temperatures yield indistinguishable $\xi_n(\phi)$ as $n$ becomes large, where $\xi_n(\phi)\propto\big(1/\pi-\rho\big(|\frac{\phi}{n}|\big)\big)\rho\big(|\frac{\phi}{n}|\big)$. Note that the state $|\boldsymbol{n}\rangle$ can be obtained from $|\boldsymbol{m}\rangle$ by exchanging the occupied states with unoccupied states (detailed at the end of Appendix.~\ref{state_of_eq}); for random $|\boldsymbol{n}\rangle$ and $|\boldsymbol{m}\rangle$, the distribution of $\theta_k$ in $\langle \boldsymbol{n} |\chi_{\{a\}} |\boldsymbol{m} \rangle$ is thus characterized by $\big(1/\pi-\rho(\theta_k)\big)\rho(\theta_k)$. For a given $|a|$, $\xi_n(\phi)$ is not sensitive to the temperature and exact $\{a\}$, except when $\xi_1(\phi)$ is sharply peaked ($\beta \rightarrow \infty$) or when the differences in indices are small.}
		\label{rho_warp}
	\end{figure}

	Several conclusions can be derived immediately:
	\begin{enumerate}
		\item $|\langle \boldsymbol{n}|\chi_{\{a_1,a_2,\cdots\}}|\boldsymbol{m} \rangle|$ depends only on the differences in indices $a_j$, since an overall shift $a_j\rightarrow a_j+\bar{a}$ only contributes an overall factor $\text{exp}\Big[\text{i}\,\bar{a}\,\big(\sum_i s_i \theta_{k_i}\big)\Big]$ to the determinant.

		\item In general, the distribution of $|\langle \boldsymbol{n} |\chi_{\{a\}} |\boldsymbol{m} \rangle|$ depends only on $|a|$ and is insensitive to the temperature or the exact index set $\{a\}$. Since $|\langle \boldsymbol{n} |\chi_{\{a\}} |\boldsymbol{m} \rangle|$ is the determinant of a matrix in which each element $\sim e^{\pm\text{i}\Delta a\,\theta}$, for a random $\{a\}$, the $\Delta a$ is typically large (since $N$ is large), while $\theta$ obeys a certain distribution that depends on temperature. Consequently, the distribution of $\pm\Delta a\,\theta$ is equivalent to one obtained by stretching the distribution of $\pm\theta$ from $(-\pi,\pi)$ to $(-\Delta a\, \pi,\Delta a\,\pi)$ and winding it around the unit circle. The effective distribution of $\pm\Delta a\,\theta$ then becomes flat and nearly irrelevant to the initial distribution of $\pm\theta$ and the exact value of $\Delta a$ (provided $\Delta a$ is large), as illustrated in Fig.~\ref{rho_warp}.

		\item For a very special $\chi_{\{a\}}$, the distribution of $|\langle \boldsymbol{n}|\chi_{\{a\}}|\boldsymbol{m} \rangle|$ can be sparse. Consider the phase $\Delta a\,\theta_k = \frac{\Delta a\,(2k-1)}{N}\pi$, if $\Delta a$ and $N$ are not relatively prime, the possible phases become restricted. For example, if $\frac{\Delta a}{N} = \frac{1}{4}$, then the phase can only be $\pm\frac{(2k-1)\pi}{4}$. Such special choices become statistically rare as $N\rightarrow\infty$.
	\end{enumerate}

	As the lower panel of Fig.~\ref{frechet_gig} shows, for a fixed $|a|$, the distribution of $|\langle \boldsymbol{n} |\chi_{\{a\}} |\boldsymbol{m} \rangle|$ is insensitive to the specific choice of index set $\{a\}$. Furthermore, as shown in Fig.~\ref{beta_}, the temperature dependence is weak. In these figures, we assume that the particle number is conserved; therefore, $|a|$ must be even for nonvanishing $\langle \boldsymbol{n} |\chi_{\{a\}} |\boldsymbol{m} \rangle$. The distribution of $\mathcal{M}_{\{a\}}$ for odd $|a|$ is also well fitted by a GIG ($p=1$) if changing particle number is allowed, as shown in Appendix.~\ref{a_23} for the case $|a|=3$. The case $|a|=2$ is special, as we discussed in Appendix.~\ref{a_23}.  In Figs.~\ref{frechet_gig},~\ref{phase_modulus} and~\ref{beta_}, we restrict to $d_{\boldsymbol{nm}} = |a|$ since this is the dominant case; for $d_{\boldsymbol{nm}} = |a|-2$ and $|a|-4$, we find that the distribution of $\mathcal{M}_{\{a\}}$ is also well fitted by a GIG ($p=1$).

	\begin{figure}[!t]
		\centering
		\includegraphics[trim=0cm 0.2cm 0.2cm 0cm, clip, width=0.494\linewidth]{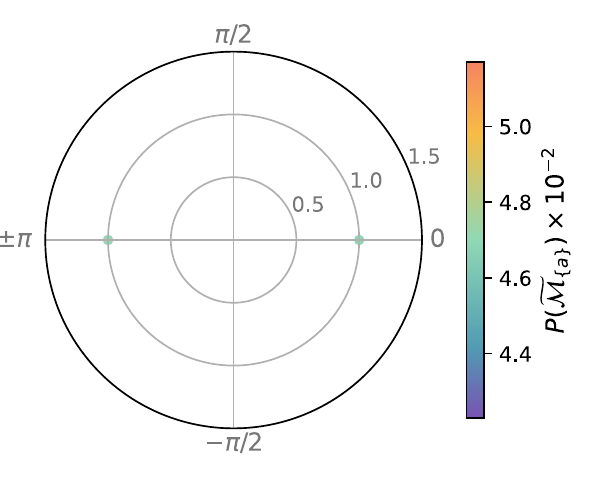}\hfill
		\includegraphics[trim=0cm 0.2cm 0.2cm 0cm, clip, width=0.494\linewidth]{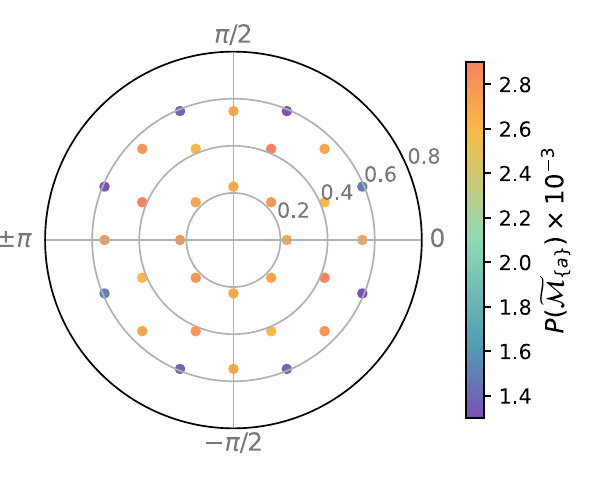}
		\caption{Distribution of $\widetilde{\mathcal{M}}_{\{a\}}=(\frac{N}{2|a|})^\frac{|a|}{2}\langle \boldsymbol{n} |\chi_{\{a\}} |\boldsymbol{m} \rangle$ in the complex plane for special $\{a\}$. $\beta=1$, $\lambda=1/2$, $N=20000$, and $10^7$ random samples for both cases. Left panel: $\{a\} = \{1,5001,10001,15001\}$; Right panel: $\{a\} = \{1,2501,5001,7501,10001,12501\}$. The color bar indicates the frequency of occurrence for each point.}
		\label{phase_modulus}
	\end{figure}
	


\Sec{Outlook}
	In this work, we have investigated the statistical properties of off-diagonal matrix elements of fermionic operators in the disorder-free Sachdev-Ye-Kitaev (SYK) model. Unlike the Lieb-Liniger model, where such matrix elements follow a Fréchet distribution, we find that in the SYK model, the distribution is well-described by a generalized inverse Gaussian (GIG) distribution for operators composed of four or more Majorana fermions. This result highlights a fundamental difference in the eigenstate thermalization behavior between integrable one-dimensional models and solvable zero-dimensional models with all-to-all interactions.
	
	Our analysis reveals that the distribution depends primarily on the size $|a|$ of the operator and is largely insensitive to the specific choice of indices or temperature, except in cases where the index differences are small or the system approaches the zero-temperature limit. The GIG fit remains robust across varying $|a|$, while the Fréchet distribution fails to capture the tails of the distribution, especially for larger $|a|$.
	
	These findings contribute to the growing understanding of the Eigenstate Thermalization Hypothesis (ETH) in solvable models and underscore the role of model dimensionality and interaction structure in determining the statistical behavior of matrix elements.
	
	In summary, our results suggest that the GIG distribution may serve as a new statistical signature for off-diagonal matrix elements in certain classes of solvable quantum many-body systems, opening avenues for further exploration at the intersection of quantum chaos, integrability, and thermalization.
	
	Several interesting directions remain for future research: 
	
	a) It would be valuable to examine whether the GIG distribution appears in other zero-dimensional or all-to-all interaction models, such as the colored SYK models or tensor models. While the disorder-free SYK model is integrable, its matrix element statistics differ from those of the Lieb-Liniger model. A systematic study of how integrability manifests in the statistics of off-diagonal matrix elements across different classes of integrable systems could provide deeper insights into quantum thermalization. 
	
	b) Although our results show weak temperature dependence, a more detailed study near critical points or in the low-temperature limit may reveal new scaling behaviors or crossover phenomena. The disorder-free SYK model exhibits weak chaos. Investigating how the GIG distribution relates to quantum chaos measures, such as out-of-time-order correlators (OTOCs) or spectral statistics, could further clarify the interplay between chaos, integrability, and thermalization. 
	

\Sec{Acknowledgments}
We are very grateful to Jia Tian for helpful discussions and suggestions. 

\Sec{Author contributions} TL conceptualized the work; SL carried out the numerical calculations; both authors contributed to the calculations and wrote the manuscript.

\bibliography{refer}

\appendix

\section{State equation and sample method\label{state_of_eq}}
	A microstate is specified by an occupation number list
	\begin{align}
		|\boldsymbol{n}\rangle = |n_1,n_2,\ldots,n_{N_F}\rangle,
	\end{align}
	where each $n_i \in \{0,1\}$. In the thermodynamic limit $N_F \to \infty$, the discrete single‑particle labels $\theta_k$ become a continuous variable $\theta \in [0,\pi]$. A macrostate is then described by a density of states $\rho(\theta)$ such that
	\begin{align}
		&N_F \rho(\theta)\,\Delta\theta = \text{number of particles in } [\theta,\theta+\Delta\theta],\nn 
		&\int_0^\pi \rho(\theta)\,d\theta = \lambda,
	\end{align}
	with $0\le \lambda\le 1$ the overall occupation rate. The number of available single‑particle levels in an interval $\Delta\theta$ is $N_F\rho_0\Delta\theta$, where $\rho_0 = 1/\pi$ is the constant density of states for the free theory. To form a macrostate, we must choose $N_F\rho(\theta)\Delta\theta$ occupied levels out of these $N_F\rho_0\Delta\theta$ possibilities. The total entropy is therefore
	\begin{align}
		S[\rho] = \sum_k \log \binom{N_F\rho_0\Delta\theta}{N_F\rho(\theta_k)\Delta\theta},
	\end{align}
	where the sum runs over intervals. In the large‑$N_F$ limit we use Stirling’s approximation:
	\begin{align}
		&\log\binom{N_F\rho_0\Delta\theta}{N_F\rho(\theta_k)\Delta\theta}\nn
		=& \log\frac{\bigl(N_F\rho_0\Delta\theta\bigr)!}{\bigl(N_F\rho(\theta_k)\Delta\theta\bigr)!\,\bigl(N_F\Delta\theta(\rho_0-\rho(\theta_k))\bigr)!} \nonumber\\
	   =& \text{const}\times\Delta\theta - N_F\Delta\theta\Bigl[\log\rho(\theta_k) + \log\rho_h(\theta_k)\Bigr],
	\end{align}
	where $\rho_h(\theta) \equiv \rho_0 - \rho(\theta)$ is the density of holes. The constant term does not affect physical quantities and is discarded. Taking the continuum limit yields the entropy functional
	\begin{align}
		S[\rho] = -N_F\int_0^\pi d\theta\,\bigl[\log\rho(\theta) + \log\rho_h(\theta)\bigr]\ .
	\end{align}
	
	The Hamiltonian of the system (in the scaling limit considered) can be expressed in terms of the densities. The quadratic part is
	\begin{align}
		H_2 = \sum_{k=1}^{N_F} \epsilon_k\Bigl(n_k - \frac12\Bigr) 
		= N_F\int_0^\pi d\theta\,\epsilon(\theta)\Bigl(\rho(\theta)-\frac12\rho_0\Bigr),
	\end{align}
	with $\epsilon(\theta)=2\cot\frac{\theta}{2}$. The quartic contribution, after proper normal ordering, takes the form
	\begin{align}
		H_4 &= \frac12\Bigl(N_F\int d\theta\,\epsilon(\theta)\bigl(\rho(\theta)-\tfrac12\rho_0\bigr)\Bigr)^2 + E_0\ .
	\end{align}
	Using the relation $H = \frac{H_4}{N_F} + \frac12\int\rho_0\,d\theta\,\epsilon(\theta)\,H_2$ (which follows from the original Hamiltonian in the large‑$N_F$ limit), we obtain
	\begin{align}
		H = \frac12 N_F\Bigl[\int_0^\pi d\theta\,\epsilon(\theta)\rho(\theta)\Bigr]^2 + \text{const}.
	\end{align}
	
	We now extremize the entropy subject to fixed energy $E = N_F\varepsilon$ and fixed particle number $\lambda$. Introducing Lagrange multipliers $\alpha$ and $\beta$, we consider
	\begin{align}
		L[\rho] &= S[\rho] + \alpha N_F\Bigl[\int_0^\pi \rho(s)\,ds - \lambda\Bigr]
		\nn
		&\hspace{10pt}+ \beta N_F\Bigl[\frac12\Bigl(\int_0^\pi d\theta\,\epsilon(\theta)\rho(\theta)\Bigr)^2 - \varepsilon\Bigr].
	\end{align}
	The functional derivative $\delta L/\delta\rho = 0$ gives
	\begin{align}
		\frac{1}{\rho(\theta)} - \frac{1}{\rho_h(\theta)} = \alpha + \beta\,\epsilon(\theta)\int_0^\pi ds\,\epsilon(s)\rho(s).
	\end{align}
	Defining
	\begin{align}
		I[\rho](\theta) \equiv \alpha + \beta\,\epsilon(\theta)\int_0^\pi ds\,\epsilon(s)\rho(s),
	\end{align}
	the equation can be rearranged to obtain
	\begin{align}
		\rho(\theta) = \frac{I[\rho](\theta) + 2\pi - \sqrt{I[\rho](\theta)^2 + 4\pi^2}}{2\pi\,I[\rho](\theta)}.
	\end{align}
	This is an integral equation for $\rho(\theta)$, since $I[\rho]$ itself depends on $\rho$. Its solution determines the equilibrium macrostate.

	For given $\beta$, the $\alpha$ is determined by the constraint
	\begin{align}
		\int_0^\pi d \theta \rho(\theta)=\lambda<1,\rho<{1\over \pi}\ .
	\end{align}
	To obtain the equilibrium density $\rho(\theta)$, one can solve the integral equation numerically. As an illustration, taking $\lambda = 1/2$ and $\beta = 1$ yields the parameters $\alpha = -4.41814141060143$ and $B = 1.2720486720980966$ (where $B$ denotes the constant $\int_0^\pi ds\,\epsilon(s)\rho(s)$ appearing in $I[\rho]$).

	For the purpose of sampling microstates consistent with a given macrostate, we follow the approach of Ref.~\cite{essler2023statisticsmatrixelementslocal}. To reduce the sampling complexity, we first divide the $\theta$‑interval $[0,\pi]$ into a suitable number of bins. Using the equilibrium density $\rho(\theta)$, we determine the expected number of fermions in each bin and then sample independently within each bin. For the evaluation in this paper, $\theta$ is divided into $100$ bins.
	
	As a concrete example, choose $N_{F} = 1000$, $\lambda = 1/2$, and $\beta = 1$, and partition $[0,\pi]$ into 20 equal bins:
	\begin{align}
		\theta[k] = \frac{k\pi}{20},\qquad k = 0,1,2,\ldots,20\ .
	\end{align}
	The number of occupied levels in each bin, obtained from the equilibrium density, is then
	\begin{align}
		N_k = \{&2,8,12,17,20,23,25,26,27,28,\nn
		&29,30,30,31,31,31,32,32,32,33\}\ .
	\end{align}
	Each bin contains $L_k = 50$ single‑particle levels. To construct a microstate, we randomly select $N_k$ levels out of the $L_k$ in each bin and occupy them; the remaining levels are left empty.
	
	To sample the matrix element $\langle \boldsymbol{n} | \chi_{\{a\}} | \boldsymbol{m} \rangle$ for fixed $\beta$ and $\lambda$, we first generate a reference microstate $|\boldsymbol{m}\rangle$ according to the above prescription. To obtain a new microstate $|\boldsymbol{n}\rangle$ from the same macrostate, we randomly pick one occupied and one unoccupied level within the same bin and swap their occupation numbers ($1\to0$ and $0\to1$). This swap preserves the number of occupied levels per bin, hence keeps the macrostate $\rho(\theta)$ unchanged, and ensures unbiased sampling across the entire $\theta$‑range.

	\begin{figure}[t!]
		\centering
		\includegraphics[trim=0.39cm 0.2cm 0cm 0cm, clip,    width=0.5\linewidth]{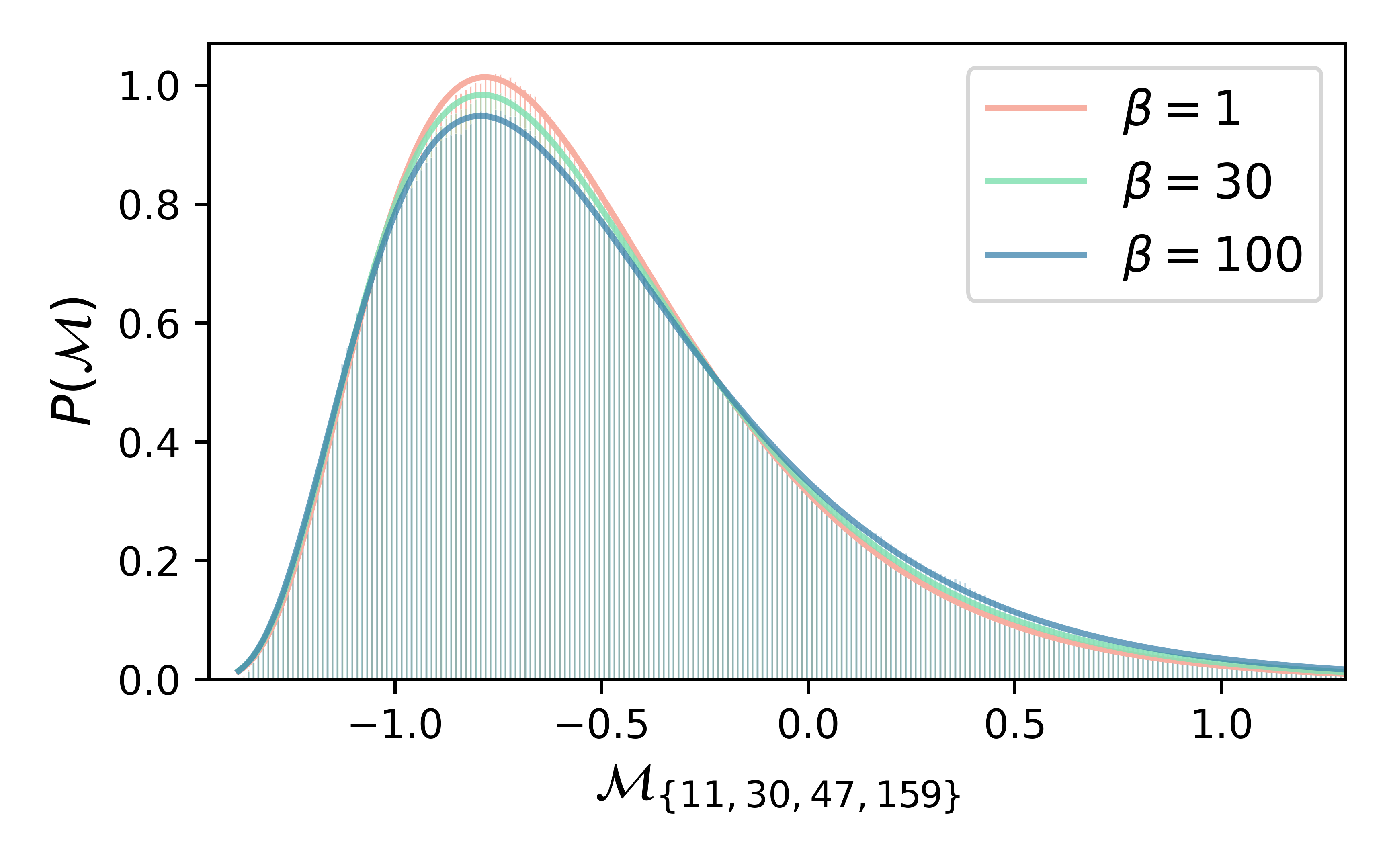}\hfill
		\includegraphics[trim=0.39cm 0.2cm 0cm 0cm, clip, width=0.5\linewidth]{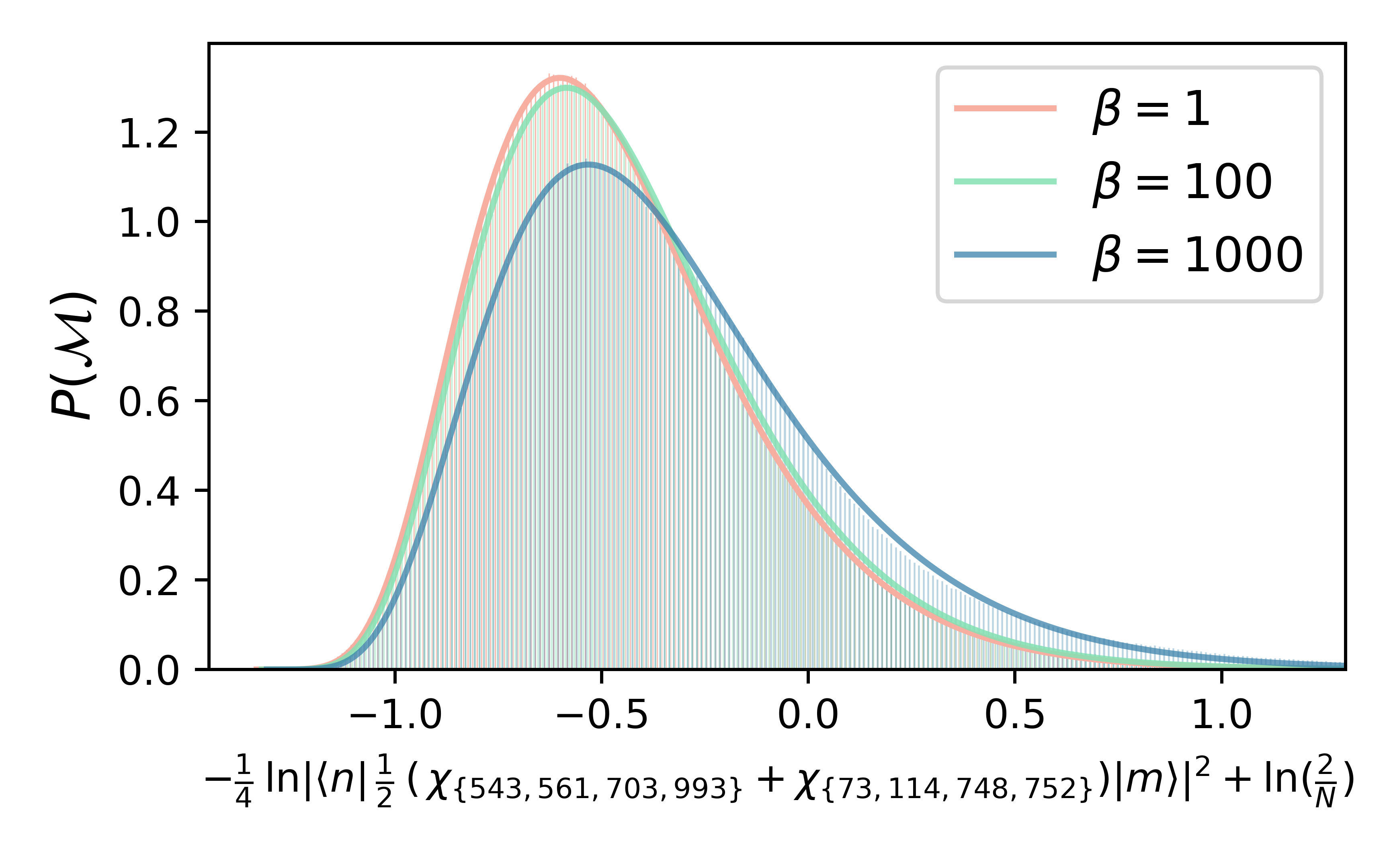}
		\caption{Distribution of $\mathcal{M}_{\{a\}}$ under different temperatures, $\lambda=1/2$, $N=20000$ and $10^7$ random samples for both cases. In the case $|a|=4$, we find that for $\delta\beta\sim100$, the temperature dependence become noticeable only when the index set $\{a\}$ includes separations $|\Delta a| \lesssim 10$. For a random choice of $\{a\}$, these curves are overlapped in general.}
		\label{beta_}
	\end{figure}

\section{Numerical Evaluation of $\langle \boldsymbol{n} |\chi_{\{a\}} |\boldsymbol{m} \rangle$\label{n_c_m}}

	Let $\boldsymbol{M}=\{p_1,\cdots,p_m\}$ and $\boldsymbol{N}=\{q_1,\cdots,q_n\}$ be the index sets of the states $|\boldsymbol{m}\rangle=f_{p_1}^\dagger \cdots f_{p_m}^\dagger|\Omega\rangle$ and $|\boldsymbol{n}\rangle=f_{q_1}^\dagger \cdots f_{q_n}^\dagger|\Omega\rangle$, respectively. To reduce computational complexity, one can map $\langle \boldsymbol{n} |\chi_{\{a\}} |\boldsymbol{m} \rangle$ to the vacuum expectation value of certain operators. Consider the map
	\begin{itemize}
		\item $f^\dagger_k\rightarrow\tilde{f}_k$ and $ f_k\rightarrow\tilde{f}^\dagger_k$, if $k\in\boldsymbol{M}$,
		\item $f^\dagger_k\rightarrow\tilde{f}^\dagger_k$ and $ f_k\rightarrow\tilde{f}_k$, if $k\notin\boldsymbol{M}$,
	\end{itemize}
	the state $|\boldsymbol{m}\rangle$ then maps into vacuum state $|\Omega\rangle$, while state $\langle\boldsymbol{n}|$ maps into
	\begin{equation*}
		(-1)^{P^\prime}\langle\tilde{\boldsymbol{n}}|\equiv (-1)^{P^\prime}\langle\Omega|\tilde{f}_{k_1}\cdots\tilde{f}_{k_{d_{\boldsymbol{nm}}}},
	\end{equation*}
	where the index set $\{k_1,\cdots,k_{d_{\boldsymbol{nm}}}\}=\boldsymbol{M}\cup\boldsymbol{N}-\boldsymbol{M}\cap\boldsymbol{N}$; the sign $(-1)^{P^\prime}$ depends on $\langle\boldsymbol{n}|$ and the order of $k_1,\cdots,k_{d_{\boldsymbol{nm}}}$. Let $\widetilde{\chi}_a$ denote the transformed $\chi_a$, we then have
	\begin{equation}
		\langle \boldsymbol{n} |\chi_{\{a\}} |\boldsymbol{m} \rangle=(-1)^{P^\prime} \langle \Omega|\tilde{f}_{k_1}\cdots\tilde{f}_{k_{d_{\boldsymbol{nm}}}}\widetilde{\chi}_{a_1}\cdots\widetilde{\chi}_{a_{|a|}} |\Omega \rangle.
	\end{equation}
	Let us define an ordered set 
	\begin{equation}
		( \mathcal{O}_1,\cdots,\mathcal{O}_{d_{\boldsymbol{nm}}+|a|})=( \tilde{f}_{k_1},\cdots,\tilde{f}_{k_{d_{\boldsymbol{nm}}}},\widetilde{\chi}_{a_1},\cdots,\widetilde{\chi}_{a_{|a|}}),
	\end{equation}
	we then reduces the evaluation of $\langle \boldsymbol{n} |\chi_{\{a\}} |\boldsymbol{m} \rangle$ to the evaluation of vacuum expectation value~\cite{Pfaffian_Method,MIZUSAKI2012219}
	\begin{equation}
			\langle\Omega|\mathcal{O}_1\cdots\mathcal{O}_{d_{\boldsymbol{nm}}+|a|}|\Omega\rangle=\text{Pf}(\hat{\mathcal{O}}),
	\end{equation}
	where $\text{Pf}(\hat{\mathcal{O}})$ is the Pfaffian~\cite{Pfaffian_Method,MIZUSAKI2012219, Pfaffian_wiki} of the matrix $\hat{\mathcal{O}}$, which is defined as
	\begin{equation}
		\hat{\mathcal{O}}_{ij}=\langle\Omega|\mathcal{O}_i\mathcal{O}_j|\Omega\rangle\quad\text{for}\quad i<j,
	\end{equation}
	and $\hat{\mathcal{O}}_{ji}=-\hat{\mathcal{O}}_{ij}$; $\text{Pf}(\hat{\mathcal{O}})=0$ when $d_{\boldsymbol{nm}}+|a|$ is odd. For a $2n\times2n$ skew-symmetric matrix $\hat{\mathcal{O}}$, the Pfaffian of $\hat{\mathcal{O}}$ is defined as
	\begin{equation}
		\operatorname {Pf} (\hat{\mathcal{O}})={\frac {1}{2^{n}n!}}\,\epsilon_{i_1j_1\cdots i_nj_n} \, \hat{\mathcal{O}}_{i_1j_1}\cdots \hat{\mathcal{O}}_{i_nj_n},
	\end{equation}
	where $\epsilon_{i_1j_1\cdots i_nj_n}$ is Levi-Civita tensor and repeated indices are summed implicitly. A special case occurs when $d_{\boldsymbol{nm}}=|a|$; in this case
	\begin{equation}
		\hat{\mathcal{O}}=\begin{pmatrix}\boldsymbol{0}&\widetilde{\mathcal{A}}\\
			-\widetilde{\mathcal{A}}^T&\widetilde{\mathcal{B}}\end{pmatrix},
	\end{equation}
	where
	\begin{equation}
		\begin{split}
		\widetilde{\mathcal{A}}_{ij}&=\langle\Omega|\tilde{f}_{k_i}\widetilde{\chi}_{a_j}|\Omega\rangle=\sqrt{\frac{2}{N}}\,e^{\text{i}s_i (a_j-1)\theta_{k_i}},\\ \widetilde{\mathcal{B}}_{ij}&=\langle\Omega|\widetilde{\chi}_{a_i}\widetilde{\chi}_{a_j}|\Omega\rangle,
		\end{split}
	\end{equation}
	where $s_i=+1$ for $k_i\in \boldsymbol{M}$ and $s_i=-1$ for $k_i\notin \boldsymbol{M}$, and we have
	\begin{equation}
		\text{Pf}(\hat{\mathcal{O}})=(-1)^{\frac{d_{\boldsymbol{nm}}(d_{\boldsymbol{nm}}-1)}{2}}\text{det}(\widetilde{\mathcal{A}}).
	\end{equation}
	This is just the Eq.~\eqref{det_A}, up to a sign depends on the order of $k_1,\cdots,k_{d_{\boldsymbol{nm}}}$.

\section{Distribution of $\mathcal{M}_{\{a\}}$ for $|a|=2, 3$ \label{a_23}}
\begin{figure}[t!]
		\centering
		\includegraphics[trim=0.39cm 0.2cm 0cm 0cm, clip,  width=0.5\linewidth]{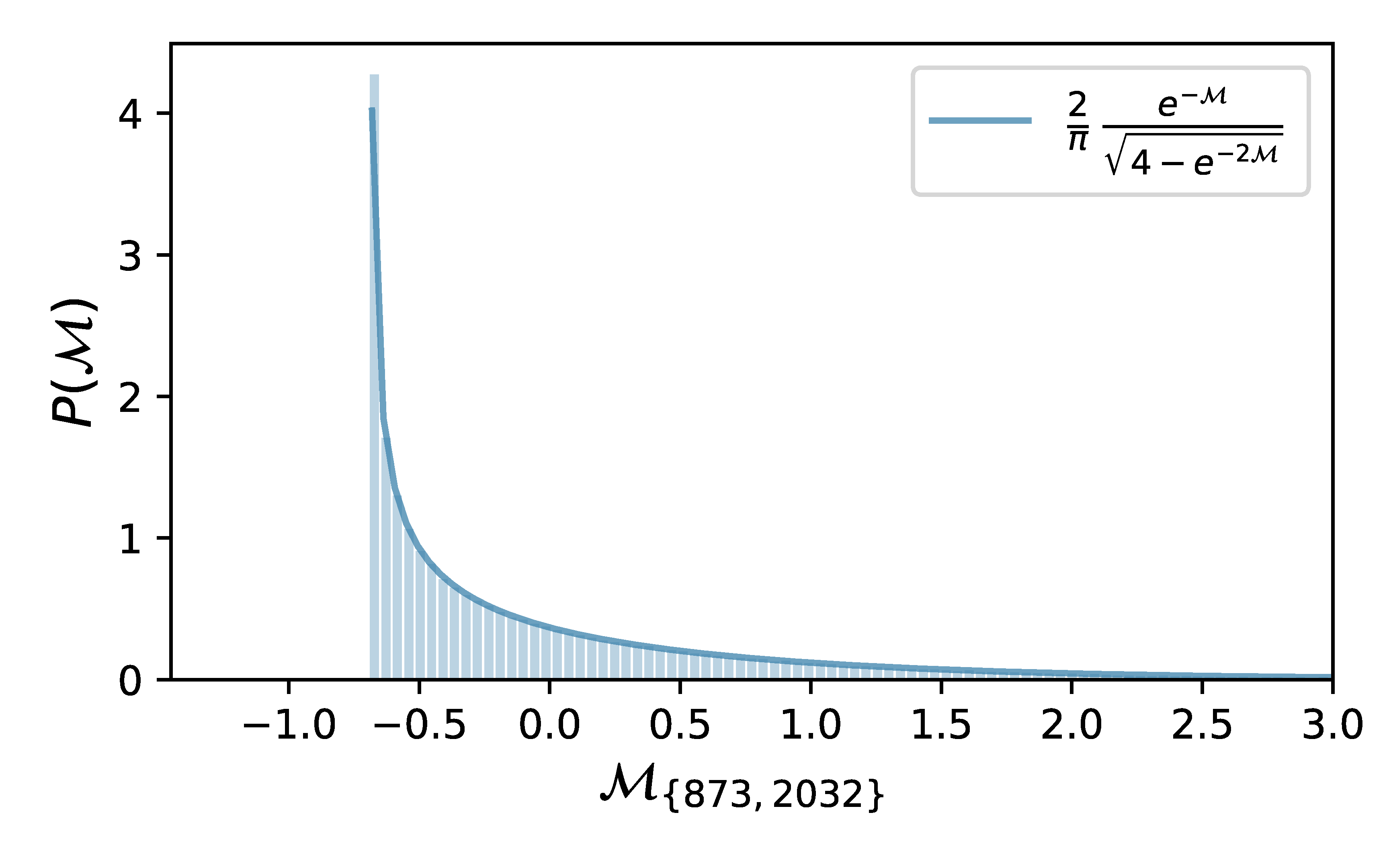}\hfill
		\includegraphics[trim=0.39cm 0.2cm 0cm 0cm, clip,  width=0.5\linewidth]{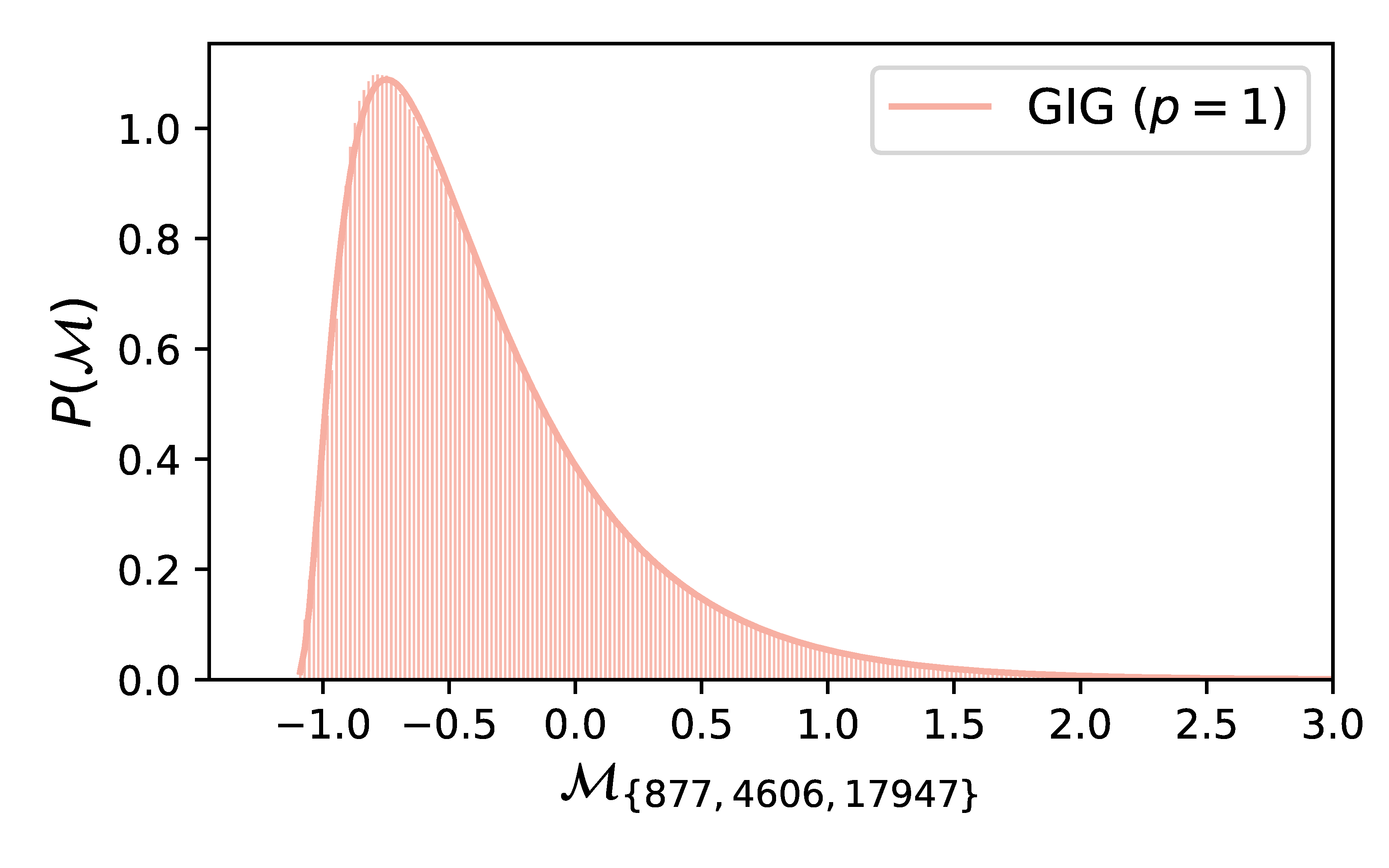}
		\caption{Distribution of $\mathcal{M}_{\{a\}}$ when $|a|=2$ (left) and $3$ (right). $\beta=1$, $\lambda=1/2$, $N=20000$, and $10^7$ random samples for both cases.}
		\label{fig_a_23}
	\end{figure}

	When $|a|=2$, the Eq.~\eqref{det_A} yields
	\begin{equation}
		\mathcal{M}_{\{a\}}=-\log|e^{\text{i}(a_1-a_2)(s_1\theta_{k_1}-s_2\theta_{k_2})}-1|.
	\end{equation}
	For general $a_1$ and $a_2$, the distribution of $\mathcal{M}_{\{a\}}$ in above can be approximated by the distribution of $-\log|e^{\text{i}\theta}-1|$, where $\theta$ is uniformly distributed from $0$ to $2\pi$, which gives $P(\mathcal{M})=\frac{2}{\pi}\frac{e^{-\mathcal{M}}}{\sqrt{4-e^{-2\mathcal{M}}}}$. As shown in Fig.~\ref{fig_a_23}, the $P(\mathcal{M})$ fitted well to the histogram. For $|a|=3$, the distribution of $\mathcal{M}_{\{a\}}$ can also be fitted well by GIG ($p=1$), with a tiny mismatch around the peak, as shown in Fig.~\ref{fig_a_23}. For smaller $|a|$, the upper bound of $|\det(\mathcal{A})|$ is easier to reach, causing the peak more asymmetric (tilt to left).

\end{document}